\documentclass[aps,
groupedaddress,showpacs,nofootinbib,prl,twocolumn]{revtex4}
\usepackage{amssymb,amsmath}
\usepackage{graphicx}
\usepackage[english]{babel}
\usepackage{graphicx}
\usepackage{dcolumn}

\usepackage{bm}
\def\nablab{\mbox{\boldmath$\nabla$}}

\def\nablabf{\mbox{\footnotesize\boldmath$\nabla$}}
\def\etab{\mbox{\boldmath$\eta$}}
\def\etabs{\mbox{\scriptsize\boldmath$\eta$}}

\def\aut#1{#1}

\def\ins#1{{\it #1}}

\def\meq#1{}
\def\meq#1{}

\def\ins#1{}

\def\comment#1{}

\def\mn#1{\marginpar[]{\scriptsize#1}}
\def\mn#1{}

\usepackage{ulem}
\usepackage{cancel}
\usepackage{color}
\def\rr#1{\textcolor{red}{#1}}

\def\mn#1{\marginpar[\tiny{\rr{#1}}]{\tiny{\rr{#1}}}}

\usepackage{hyperref}

\newcommand{\be}{\begin{equation}}\newcommand{\ee}{\end{equation}}
\newcommand{\bea}{\begin{eqnarray}}\newcommand{\eea}{\end{eqnarray}}
\newcommand{\beaa}{\begin{eqnarray}}\newcommand{\eeaa}{\end{eqnarray}}
\newcommand{\ba}{\begin{array}}\newcommand{\ea}{\end{array}}
\newcommand{\bit}{\begin{itemize}}\newcommand{\eit}{\end{itemize}}
\newcommand{\ben}{\begin{enumerate}}\newcommand{\een}{\end{enumerate}}

 \newcommand{\sfrac}[2]{\raisebox{0.095ex}{\scriptsize${\frac{#1}{#2}}$}}

\def\bra#1{\mathinner{\langle{#1}|}}	\def\ket#1{\mathinner{|{#1}\rangle}}

									\def\be{\begin{equation}}
\def\ee{\end{equation}}				\def\bea{\begin{eqnarray}}				\def\eea{\end{eqnarray}}
\def\bear{\begin{array}}				\def\eear{\end{array}}

												\def\x5{x^{5}}

\begin{document}

\title{Green function of
the double fractional Fokker-Planck equation --\\ Path integral and stochastic differential equations}

\comment{
\title{Fractional
Fokker-Planck Equations,
 Stochastic Differential Equations, \\
and Path Integrals
for L\'evy Random Walks
}
}

\comment{
\author{H. Ruby}
\email{hruby@live.de}
\affiliation{ICRANeT Piazzale della Repubblica, 10 -65122, Pescara, Italy}
}

\author{H. Kleinert}
\email{h.k@fu-berlin.de}


\affiliation{Institut f{\"u}r Theoretische Physik, Freie Universit\"at Berlin, 14195 Berlin, Germany}
\affiliation{ICRANeT Piazzale della Repubblica, 10 -65122, Pescara, Italy}

\author{V. Zatloukal}
\affiliation{Faculty of Nuclear Sciences and Physical Engineering, Czech Technical University in Prague, B\v{r}ehov\'{a} 7, 115 19 Praha 1, Czech Republic}
\affiliation{Max Planck Institute for the History of Science, Boltzmannstrasse 22, 14195 Berlin, Germany}


\vspace{2mm}
\def\sch{Schr\"odinger}
\def\comment#1{}

\begin{abstract}
The statistics of rare events,
the so-called {\it Black-Swan Events},
is goverened by non-Gaussian distributions 
with heavy power-like tails.
We calculate the Green functions of the associated Fokker-Planck equations
and solve the related
stochastic differential equations.
We also discuss the subject
in the framework of path integration.
\end{abstract}




\maketitle

\section{Introduction}

Gaussian random walks
prove to be a natural and rather universal
starting point for
many stochastic processes.
In fact, the famous {\it central-limit theorem\/}
shows that
many independent random movements
of finite variance $\sigma^2=\langle x^2\rangle$
always pile up to display
a Gaussian distribution
 \cite{BP}.
In particular, Gaussian random walks constitute
the basis of the most important tool
in the theory of financial markets,
the Black-Scholes option price theory \cite{BLS} (Nobel Prize 1997),
by which a portfolio of assets
is hoped to remain steadily growing through hedging \cite{PIR2}.

However, since the last stock market crash and the
still ongoing
financial crisis
it has become clear
that 
distributions
which describe realistically the behaviour of financial markets
belong to a more general universality class,
the so-called  L\'evy stable distribution \cite{PREIS,MANST,PODOB}.
They result from a
sum of
random movements
of infinite variance \cite{remn}, and 
account
for the fact that
rare events, the so-called Black-Swan Events \cite{TAL}, which
initiate crashes,
are much more frequent
than in Gaussian distributions.
These are events in
the so-called  L\'evy tails
$\propto 1/|x|^{1+\lambda}$ of the distributions, 
whose
description
is based on a generalized Hamiltonian  \cite{PIR}:
\begin{eqnarray}
H(p)={\rm const}\,(p^2)^{\lambda/2}.
\label{@HAM}\end{eqnarray}
Such tail-events are present
in many physical situations, e.g., in velocity distributions of many body systems with long-range forces 
\cite{RUFFO},
in the
self-similar
 distribution of matter in the universe \cite{EI,VS,DU}, 
 and
in the distributions
of windgusts 
\cite{PEINKE}
and
earthquakes
\cite{SEISM},
with often catastrophic consequences.

Distributions with L\'evy tails are a consequence
of rather general maximal entropy assumptions \cite{GMTS}. 
In the limit $\lambda\rightarrow 2$, the  L\'evy  distributions
reduce to Gaussian  distributions. 

The simplest L\'evy-type random walk
is described by the 
stochastic differetial equation 
 of the
Langevin type
\begin{eqnarray}
\frac{ d}{ds}x(s)\equiv\dot x(s)=\eta(s),
\label{@19}\end{eqnarray}
where $\eta(s)$ is a noise variable
as a function of a pseudotime $s$ 
with zero expectation value and a probability distribution
characterized by a parameter $\lambda$
 \cite{REM2}:
\begin{eqnarray}\!\!\!\!P[\eta]
\!\equiv\!  e^{-\int ds \tilde H(\eta)}\!=\!\!
\int{\cal D} p\exp\left\{\int ds\,\left[ i
p
\eta
-(p^2)^{\lambda/2}\right]
\right\}\!\!.
\label{@21}\end{eqnarray}
Using this we may solve
the
stochastic differential equation (\ref{@19})
in which the noise $\eta(s)$
has nonzero 
 correlation functions for even $n=2,4,6,\dots~$:
\begin{eqnarray}
\comment{
\langle \eta(s_1)
\eta(s_{2})\rangle& \equiv&
\int {\cal D}\eta\,\eta(s_1)
\eta(s_{2})P[\eta]
, \\
&\vdots&\nonumber \\
}
\langle \eta(s_1)\dots
\eta(s_{n})\rangle &\equiv&
\int {\cal D}\eta\,\eta(s_1)\dots
\eta(s_{n})P[\eta].
\label{@FPFR}\end{eqnarray}
For $\lambda=2$,
the distribution is Gaussian
and $\eta(s)$ is a standard white noise variable.
If we solve  (\ref{@19}) in $D$ dimensions with an initial condition 
${\bf x}(0)={\bf 0}$, the variable ${\bf x}(s)$
has  a distribution 
\begin{eqnarray}
P_{\rm G}({\bf x},s)=
{(4\pi s)^{-D/2}}e^{-{\bf x}^2 /4s}.
\label{@GDD}\end{eqnarray}
This distribution is the Green function of the Fokker-Planck equation
\begin{eqnarray}
(\partial_s+
\hat{\bf p}^2
)P_G({\bf x},s)=\delta(s)\delta^{(D)}({\bf x})
,
\label{@10}\end{eqnarray}
where
$\hat {\bf p}\equiv i\partial_{\bf x}\equiv i \nablab$.
For $\lambda\neq 2$, the distribution is non-Gaussian and it solves the
{\it fractional
Fokker-Planck equation\/}
\begin{eqnarray}
[\partial_s+
(\hat{\bf p}^2)^{\lambda/2} 
]P({\bf x},s)=\delta(s)\delta^{(D)}({\bf x})
.
\label{@10G}\end{eqnarray}
A solution of this equation that evolves from the $\delta$-function is
\begin{equation}
P({\bf x},s)=e^{-s (\hat{\bf p}^2)^{\lambda/2}} \delta^{(D)}({\bf x}),
\end{equation}
and for $s=1$ it coincides with the noise probability,
\begin{equation}
P({\bf x},1)|_{{\bf x}={\bm \eta}}=P({\bm \eta}) = \int \frac{d^D p}{(2\pi)^D}  e^{i{\bf p}{\bm \eta}-({\bf p}^2)^{\lambda/2}} .
\end{equation}
Applications of the fractional Fokker-Planck equation 
are numerous in non-Brownian diffusion processes. 
These are observed in 
chaotic systems and in the fluid dynamics of rheology and biology. 
See \cite{ShlZas,PekSzn} for an overview.
The mathematics of Eq. (\ref{@10G}) 
with variable diffusion coefficient is in \cite{Sro}.

The fractional Fokker-Planck  equation 
(\ref{@10G})
can be generalized further
to the
 {\it double
fractional Fokker-Planck equation\/}
\begin{eqnarray}
[\hat{p}_4^{1-\gamma}+D_\lambda(\hat{\bf p}^2)^{\lambda/2} 
]P({\bf x},t)=\delta(t)\delta^{(D)}({\bf x})
,
\label{@10G2}\end{eqnarray}
where
 $\hat{p}_4\equiv \partial_t$,
$\hat {\bf p}\equiv i\partial_{\bf x}\equiv i \nablab$
and  a parameter has been allowed for that is the analogue of the 
diffusion constant $D$ in the ordinary diffusion process
\cite{REMMK}.

We should explain the physical origin
of the fractional powers in the space and time derivatives
of the above equation.
Such powers occur naturally in many-particle systems 
if the interaction strength or the range becomes very large. 
As long as the interaction strength is small and the range is short, 
such systems are described by a second-quantized field theory 
with a free-particle action
\begin{eqnarray}
{\cal A}_0 \!=\! \int \!\! dt d^3x \psi^\dagger({\bf x},t) (i\partial_t \!+\! \hbar^2 \nabla^2/2m \!-\! V({\bf x})) \psi({\bf x},t) ,
\label{@}\end{eqnarray}
and an interaction 
of the type
\begin{eqnarray}
{\cal A}_{\rm int}=\frac{g}{4!}\int dt d^3x(\psi^\dagger\psi)^2 \ .
\label{@}\end{eqnarray}
The partition function 
can be calculated from the functional integral
\begin{eqnarray}
Z=\oint {\cal D}\psi{\cal D}\psi^\dagger e^{i({\cal A}_0+{\cal A}_{\rm int})/\hbar}.
\label{@}\end{eqnarray} 
A  perturbation expansion 
 leads to an effective action
in the form of a 
power series  of $g\Psi^\dagger\Psi $, where $\Psi=\langle \psi\rangle$ are the expectation values of the field.
This series is divergent 
and must be resummed. For large interaction strength
$g$,
this produces 
anomalous power behaviors 
in the field strength as well as in the momenta \cite{KS,HKN}. 
The free-field part of the effective 
action leads to a field equation of the fractional Fokker-Planck 
or Schr\"odinger type,
 in which momentum and energy
appear with powers different from $\lambda=2$ and $\gamma=0$, respectively.

In addition, equations of the type (\ref{@10G2}) 
are known to govern various different phenomena. In chaotic systems, for example,  they describe
 anomalous diffusion processes with memory (time non-locality) \cite{MEER,ZAS}.
In fact, the fractional time derivatives also arise as the infinitesimal generators of coarse grained time evolutions \cite{Hil}, or they can be derived from a random walk model when the mean waiting time of the walker diverges 
\cite{BMK}.

It is the purpose of this note to 
 calculate the Green functions of general fractional
Fokker-Planck equation (\ref{@10G2})
and specify the  path integrals 
solved by them \cite{PI,Feynman}.

%

\section{Double fractional Fokker-Planck equation}

\begin{figure}
\includegraphics[scale=1]{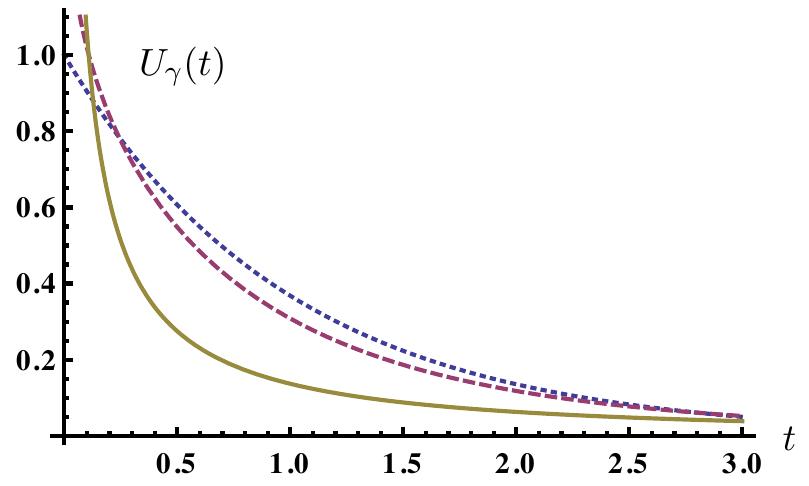}
\caption{(Color online) The function $U_\gamma(t)$ for $\hat{H}=1$, and various values of $\gamma$. Dotted (blue) curve: $\gamma=0$, standard exponential function; Dashed (red) curve: $\gamma=0.1$; Solid (yellow) curve: $\gamma=0.5$.}
\label{figML}
\end{figure}

A convenient definition of the fractional derivatives
uses the same formula
as in the dimensional 
continuation of Feynman diagrams \cite{METZ,METZz},
\begin{equation} \label{@9a}
(\hat{\bf p}^ 2)^{\lambda/2}=\Gamma[-\lambda/2]^{-1}\int d\sigma \sigma^{-\lambda/2-1}
e^{\sigma \hat{\bf p}^ 2}.
\end{equation}
\comment{
$\hat p_4^{1-\gamma}$ and $(\hat {\bf p}^2)^{\lambda/2}$
$\hat p_4^{1-\gamma}f(t)
\equiv \Gamma^{-1}[1-\gamma]\int_t^\infty dt'(t-t'+i\epsilon)^{-2+\gamma}
f(t')$ \cite{RAI}.
The $D$-dimensonal operaton
$\hat p_4^{\lambda/2}f({\bf x})$}%
The solution of (\ref{@10G2}) can be written formally as
\begin{equation}
P({\bf x},t) =
[(\hat{p}_4+\epsilon)^{1-\gamma}+D_\lambda(\hat{\bf p}^2)^{\lambda/2}]^{-1}
\delta(t)\delta^{(D)}({\bf x})
,
\label{@10G3}\end{equation}
where infinitesimal $\epsilon > 0$ ensures forward-in-time nature of the Green function, and its explicit appearance will be suppressed from now on.
Using the representation $\delta(t) = \int_{-\infty}^{+\infty} \frac{dE}{2\pi} e^{-i E t}$, we arrive at
\begin{equation} \label{@P1}
P({\bf x},t) =
\int \frac{dE}{2\pi} \frac{e^{-i E t}}{(-iE)^{1-\gamma} + D_\lambda(\hat{\bf p}^2)^{\lambda/2}} \delta^{(D)}({\bf x}).
\end{equation}
Now we expand the fraction into a geometric series, and integrate term by term using the formula \cite{GRAD} 
\begin{equation}
\int_{-\infty}^{+\infty} \frac{dE}{2\pi} \frac{e^{-i E t}}{(-iE+\epsilon)^{(1-\gamma)(n+1)}} = \frac{\theta(t) t^{n(1-\gamma)-\gamma}}{\Gamma[(1-\gamma)(n+1)]},
\end{equation}
where $\theta(t)$ is the Heaviside step function.
The result can be cast as
\begin{equation}
P({\bf x},t) = \theta(t) t^{-\gamma} E_{1-\gamma,1-\gamma}[-t^{1-\gamma} D_\lambda(\hat{\bf p}^2)^{\lambda/2}] \delta^{(D)}({\bf x}),
\end{equation}
where $E_{\alpha,\beta}(z) = \sum_{n=0}^\infty \frac{z^n}{\Gamma(\alpha n + \beta)}$ is the {\it Mittag-Leffler} function \cite{Erdelyi,Haubold}. 
This can be interpreted by writing
\begin{equation}
P({\bf x},t) = \bra{\bf x} \hat{U}_\gamma(t) \ket{\bf 0},
\end{equation}
with the $\gamma$-deformed evolution $\hat{U}_\gamma$ defined by
\begin{equation}
\hat{U}_\gamma(t) = \theta(t) t^{-\gamma} E_{1-\gamma,1-\gamma}(-t^{1-\gamma} \hat{H}),
\end{equation}
with $ \hat{H} \equiv D_\lambda(\hat{\bf p}^2)^{\lambda/2}$ \cite{REMHamiltonian}
 (See Fig. \ref{figML}.). 
The occurrence of the Mittag-Leffler function in solutions 
of the time-fractional Fokker-Planck equation has been noted previously, 
for example, in the review article \cite{REMMK}.

\begin{figure}
\includegraphics[scale=1]{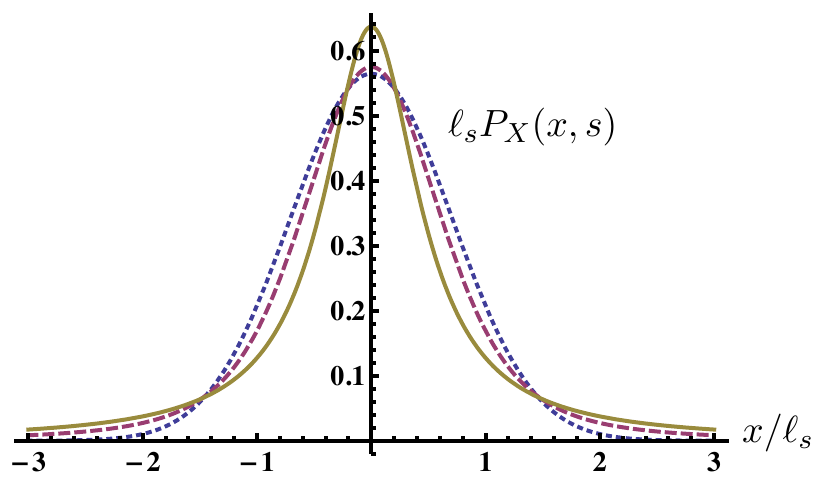}
\caption{(Color online) Dotted (blue) curve: $\lambda=2$, standard Gaussian distribution; Dashed (red) curve: $\lambda=1.5$; Solid (yellow) curve: $\lambda=1$, Cauchy-Lorenz distribution. The length scale is $\ell_s = 2(D_\lambda s)^{1/\lambda}$.}
\label{figPX}
\end{figure}

For $\gamma=0$, the equation (\ref{@10G2}) reduces to a single (space) fractional Fokker-Planck equation
\begin{equation} \label{@SF1}
[\hat{p}_4 + D_\lambda (\hat{\bf p}^2)^{\lambda/2}] P({\bf x},t) = \delta^{(D)}({\bf x})\delta(t),
\end{equation}
the Mittag-Leffler function reduces to $E_{1,1}(z) = \exp(z)$, and the evolution operator recovers its standard form $\hat{U}_0(t) = \theta(t) \exp(-t \hat{H})$. The solution, which we shall denote by
$P_X({\bf x},t)$ for a more specific reference, is the
{\it multivariate L\'{e}vy stable distribution} \cite{Nolan}:
\begin{equation}
P_X({\bf x},t) = \int \frac{d^D p}{(2\pi)^D}
e^{-t D_\lambda ({\bf p}^2)^{\lambda/2}} e^{-i {\bf p}{\bf x}}.
\label{@PX1} \end{equation}
For $\lambda =2$, it reduces to  the
standard quantum mechanical Gaussian expression (\ref{@GDD}).
For $\lambda =1$,
the result is
\begin{eqnarray}\!\!\!\!
P_X({\bf x},t) = \frac{[\Gamma(D/2+1/2)/\pi^{(D+1)/2}]D_\lambda t}{[(D_\lambda t)^2+|{\bf x}|^2]^{D/2+1/2}},
\label{@13A}\end{eqnarray}
which is the Cauchy-Lorentz distribution function. In Fig. \ref{figPX}, we plot $P_X$ in $D=1$ dimension for $\lambda=1,1.5,2$.

In the Appendix we provide various useful representations of $P_X({\bf x},t)$.
At this place it is worth
mentioning that
this probability
can be written as
a superposition 
\comment{
$
\int_0^\infty( d\sigma/\sigma)
 f_\lambda(\sigma t ^{-2/\lambda})P_{\rm G}(\sigma,{\bf x})$
}
of Gaussian
distributions
$ 
P_{\rm G}(\sigma,{\bf x})=
{(4\pi \sigma)^{-D/2}}e^{-{\bf x}^2 /4\sigma}$
\comment{
\begin{eqnarray}
f_\lambda(\sigma)=
S_D\sum_{n=1}^\infty
\frac{(-1) ^ n{\sigma^{-n \lambda/2}}}{(n+1)!\Gamma(D-1-n \lambda/2)}D_\lambda^{n/\lambda},
\label{@}\end{eqnarray}
where $S_D=2\pi^{D/2}/\Gamma(D/2)$ is the surface of a sphere in $D$ dimensions.
}
to be specified in Eq.~(\ref{@52}).

\subsection{Smeared-time representation, and relation between physical time $t$ and pseudotime $s$}

If we use in (\ref{@P1}) the Schwinger's formula $1/A = \int_0^\infty \!\!ds~ e^{-s A}$, we can express $P({\bf x},t)$ as an integral
\begin{equation}
P({\bf x},t) = \int_{0}^{\infty} \!\!\!ds P_X({\bf x},s) P_T(t,s) ,
\label{@PXT1}\end{equation} 
where $P_X$ solves the {\it space-fractional} diffusion equation (\ref{@SF1}), with $t \equiv s$, and $P_T$ solves the {\it time-fractional} equation
\begin{equation} \label{@TF1}
[\partial_s + \hat{p}_4^{1-\gamma}] P_T(t,s) = \delta(t)\delta(s),
\end{equation}
which encodes the relation between the {\it pseudotime} $s$ and the physical time $t$. 
The factorized ansatz (\ref{@PXT1}) has been used previously in \cite{BarSil} to solve the time-fractional Fokker-Planck equation.

For $\gamma=0$, $P_T(t,s) = \delta(t-s)$, and (\ref{@PXT1}) reduces to $P({\bf x},t)=P_X({\bf x},t)$. 

\begin{figure}
\includegraphics[scale=1]{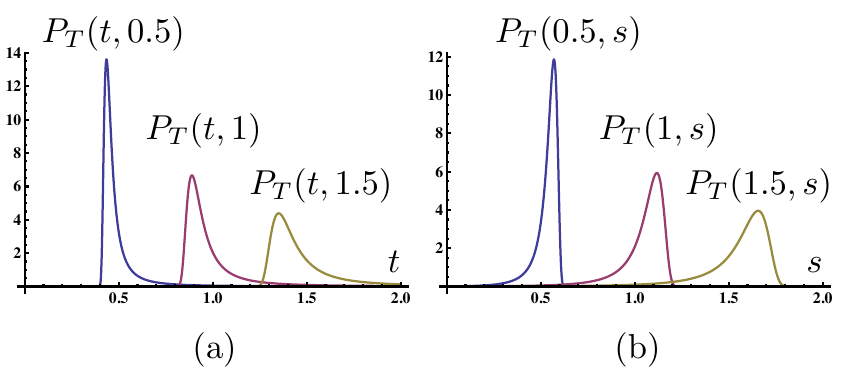}
\caption{(Color online) (a) $P_T(t,s)$ as a distribution of $t$ with increasing values of the pseudotime $s=0.5,1,1.5$. (b) $P_T(t,s)$ as a distribution of $s$ with increasing values of the real time $t=0.5,1,1.5$. In both cases $\gamma=0.03$.}
\label{figPT}
\end{figure}

For $\gamma > 0$, we obtain an {\it asymmetric L\'{e}vy stable distribution} \cite{REMstable}
\begin{equation} \label{@PT}
P_T(t,s) = \int_{-\infty}^{\infty} \frac{dE}{2\pi} e^{-s (-i E)^{1-\gamma}} e^{-i E t}.
\end{equation}
An important feature is that $P_T(t,s)$ vanishes for $t<0$. This can be seen by placing the branch cut of a multivalued function $z^{1-\gamma}$ along the negative real axis, and 
calculating (\ref{@PT}) as a complex integral with contour that follows the real axis, and closes in the upper half-plane. See Fig. \ref{figPT} (a) where $P_T$ is plotted as a function of $t$ for the case $\gamma=0.03$, and various values of $s$.

It is illustrative to view formula (\ref{@PXT1}) as a smearing of the distribution $P_X({\bf x},s)$ around the time position $t$, defined by the probability density function $P_T(t,s)$. For this purpose we plot in Fig. \ref{figPT} (b) $P_T(t,s)$ as a function of $s$, with parameter $t$ describing the position of the peak in the probability distribution.

\begin{figure}
\includegraphics[scale=1]{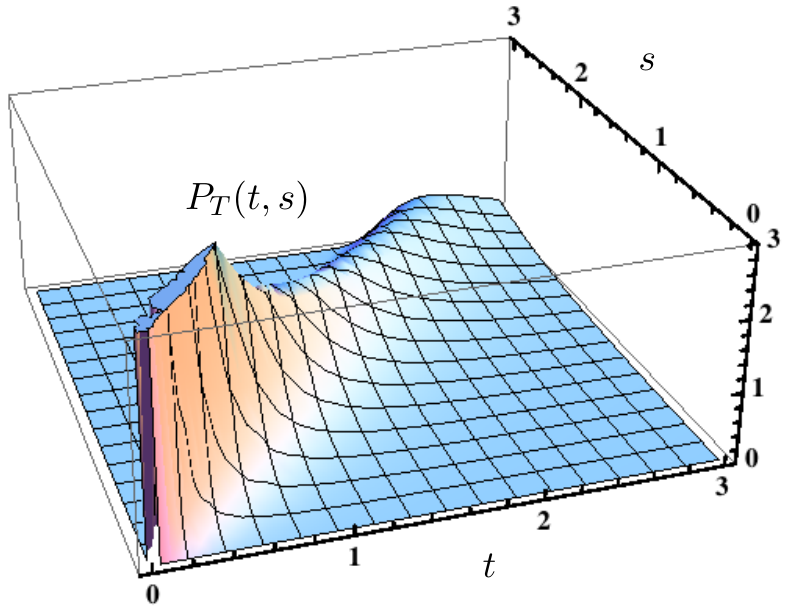}
\caption{(Color online) $P_T(t,s)$ as a function of both $t$ and $s$. Here $\gamma=0.1$.}
\label{figPT3}
\end{figure}

The two plots in Fig. (\ref{figPT}) are related through the formula 
\begin{equation}
P_T(t,s) = (C/t) P_T(C,C^{1-\gamma} t^{\gamma-1} s),
\end{equation}
which can be deduced from (\ref{@PT}) by a simple change of the integration variable $E \rightarrow (C/t) E$. Here $C$ is an arbitrary constant. The function $P_T(t,s)$ as a function of two variables is shown in Figure (\ref{figPT3}).

When $\gamma=0$, $P_T(t,s) = \delta(t-s)$ is concentrated at the point $t$, i.e., there is no smearing. For increasing $\gamma$ the peak around $t$ broadens, which can be accounted for by derivatives of the $\delta$-function. The action of $P_T$ on a test function $f(s)$ is
\begin{equation}
\int_0^\infty \!\!\!ds~ P_T(t,s) f(s)
= \sum_{n=0}^\infty \frac{f^{(n)}(t)}{n!} \int_0^\infty \!\!\!ds~ P_T(t,s) (s-t)^n .
\label{@}\end{equation}
We represent $f^{(n)}(t) = (-1)^n \int d\tau \delta^{(n)}(\tau-t) f(\tau)$, and calculate
\begin{equation}
\int_0^\infty \!\!\!\!\!ds P_T(t,s) s^k
\!=\!\! \int\! \frac{dE}{2\pi} \frac{e^{-i E t} k!}{(-i E)^{(1-\gamma)(k+1)}}
\!=\! \frac{k! \theta(t) t^{(1-\gamma)k-\gamma}}{\Gamma[(1\!-\!\gamma)(k\!+\!1)]}
\label{@}\end{equation}
to find that
\begin{equation}
P_T(t,s) = \sum_{n=0}^\infty \frac{t^n}{n!} c_n(t) \delta^{(n)}(s-t) ,
\label{@}\end{equation}
where
\begin{equation}
c_n(t) = \sum_{k=0}^n {n \choose k} (-1)^k \frac{k! \theta(t) t^{-\gamma(k+1)}}{\Gamma[(1-\gamma)(k+1)]} .
\label{@}\end{equation}
In view of these relations, the equation (\ref{@PXT1}) translates into
\begin{equation}
P({\bf x},t) = \sum_{n=0}^\infty \frac{(-t)^n}{n!} c_n(t) \partial_t^n P_X({\bf x},t).
\end{equation}
One can easily verify that for $\gamma=0$, $c_n = \delta_{n 0}$, and $P({\bf x},t) = P_X({\bf x},t)$.

\subsection{Fox $H$-function representation of the Green function}

\begin{figure}
\includegraphics[scale=1]{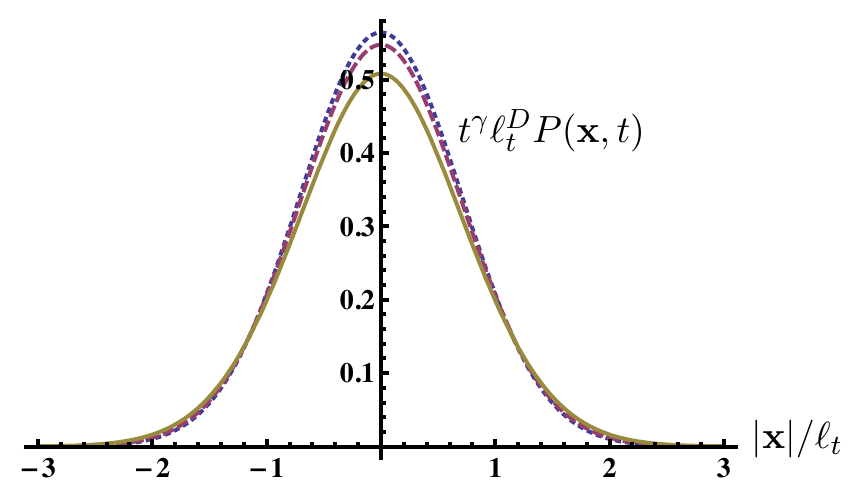}
\caption{(Color online) In all cases $\lambda=2$. Dotted (blue) curve: $\gamma=0$, standard Gaussian distribution; Dashed (red) curve: $\gamma=0.03$; Solid (yellow) curve: $\gamma=0.1$.}
\label{figPXT}
\end{figure}

Solution of the double fractional equation (\ref{@10G2}) has been obtained previously in terms of the {\it Fox H-function} \cite{DUAN}. We derive the same result starting from formula (\ref{@PXT1}), where we consider the representation (\ref{@PX2}) of $P_X({\bf x},s)$. Integration over the pseudotime $s$ can be performed, followed by the $E$ integration, that yields
\begin{equation}
P({\bf x},t) = 
\!\!\frac{t^{-\gamma}}{\pi^ {D/2}|{\bf x}|^{D}}H^{2,1}_{2,3}
\left(\!
\left[\frac{|{\bf x}|}{
\ell_t}\right]^ \lambda
\bigg |
^{(1,1);(1-\gamma,1-\gamma)}
_{(1,1),(D/2,\lambda/2);(1,\lambda/2)}\!
\right)\!.
\label{@11}
\end{equation}
Here $\ell_t\equiv 2 (D_\lambda t^{1-\gamma})^{1/\lambda}$ is a $t$-dependent
length scale, and $H^{2,1}_{2,3}$ is the Fox $H$-function \cite{FH,Mathai}, defined by the contour integral
\begin{equation}
\frac{P({\bf x},t)|{\bf x}|^{D}}{t^{-\gamma} \pi^ {-D/2}} \!=
\!\!\!
\int_{\cal C} \frac{dz}{2\pi i} \frac{\Gamma(1+z) \Gamma(\frac{D}{2}+\frac{\lambda}{2}z) \Gamma(-z)}{\Gamma(-\frac{\lambda}{2}z) \Gamma(1\!-\!\gamma+(1\!-\!\gamma)z)}
\!\!\left[
\frac{|{\bf x}|^{\lambda}}{
\ell_t^\lambda} \right]^{-z}\!\!\!\!,
\label{@11I}\end{equation}
where the contour ${\cal C}$ runs from $- i\infty$ to $+ i\infty$. In Fig. \ref{figPXT} we show how values of $\gamma>0$ modify the Gaussian distribution (for which $\lambda=2$, $\gamma=0$).

The large-$|{\bf x}|$ asymptotics of (\ref{@11}) is governed by the pole of the integrand at $z=1$:
\begin{eqnarray}
t^\gamma|{\bf x}|^{D}
P({\bf x},t) 
\overset{|{\bf x}|\rightarrow\infty}
{\approx} 
\frac{ \ell_t^{\lambda} }{|{\bf x}|^{\lambda}}
  \frac{-\Gamma(\frac{D+\lambda}{2})}
{\pi^{D/2}\Gamma(2-2\gamma) \Gamma(-\frac{\lambda}{2})} .
\end{eqnarray}

%
\comment{
The general properties of fractional differential operators
are discussed in \cite{LEVY}.
}

\comment{
Their
treatment in terms of Gaussian random walks
has been developed in \cite{Feynman}, and its generalization
to  L\'evy walks
was discussed in \cite{PI}.
}

\begin{figure}
\includegraphics[scale=0.9]{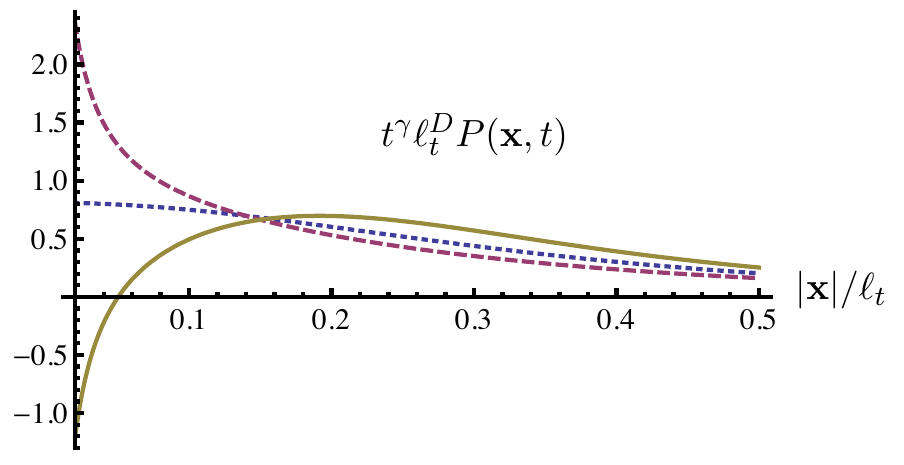}
\caption{(Color online) Dotted (blue) curve: $\gamma=0, \lambda=1$, Cauchy-Lorentz distribution; Dashed (red) curve: $\gamma=0.1, \lambda=1$; Solid (yellow) curve: $\gamma=-0.1, \lambda=1$, assumes negative values. Here D=3.}
\label{fig2}
\end{figure}

Analysis of the small-$|{\bf x}|$ behavior is more subtle due to a richer pole structure of the integrand in (\ref{@11I}) (see \cite{Kilbas}). If we assume only simple poles, we can extract the leading behavior
\begin{eqnarray}
t^\gamma P({\bf x},t) \overset{|{\bf x}|\rightarrow 0}{\approx} 
\left\{
\begin{array}{ll}
A(t) + B(t) |{\bf x}|^{2\lambda-D} ,& 2\lambda-D<2 \\
A(t) + {\cal O}[|{\bf x}|^2](t) ,& 2\lambda-D>2
\end{array}
\right.,
\end{eqnarray}
with
\begin{eqnarray}\!\!\!\!\!\!\!\!\!
A(t)\!& =&\! \frac{\pi^{1-D/2} \ell_t^{-D} 2/\lambda}{\sin(\pi\frac{D}{\lambda}) \Gamma(\frac{D}{2})\Gamma[\frac{(1-\gamma)(\lambda-D)}{\lambda}]} ,
\\  \!\!\!\!\!\!\!\!\!
B(t)\!& =&\! -\frac{\pi^{-D/2} \Gamma(\frac{D}{2}-\lambda)}{\Gamma(\lambda) \Gamma(\gamma-1) \ell_t^{2\lambda}} .
\end{eqnarray}
In particular, for $2\lambda < D$ the value of $P({\bf x},t)$ tends to either $+\infty$ or $-\infty$ as $|{\bf x}|\rightarrow 0$.
See Fig. \ref{fig2}.

\section{Path-integral formulation}

We note that the probability (\ref{@10G3})
may
 be calculated from the
 {\it doubly fractional canonical path integral\/}
over
 fluctuating orbits $t(s), {\bf x}(s)$
 $p_4(s),{\bf p}(s)$
viewed as functions of some
pseudotime $s$ \cite{REM1}:
\begin{eqnarray}
\{  {\bf x}_b  t_b s_b| {\bf x}_a t_s s_a\}=\int
{\cal D} {\bf x}{\cal D}t
 {\cal D} {\bf p}{\cal D}p_4
e^ {{\cal A}},
\label{@CFP}\end{eqnarray}
with ${\cal A}$ being the euclidean action
of the paths $t(s),{\bf  x}(s)
$:
\begin{eqnarray}
{\cal A}=\int ds[i( {\bf p} {\bf x}'-i
p_4t')-{\cal H}({\bf p},p_4)].
\label{@}\end{eqnarray}
Here
 $t'(s)\equiv dt(s)/ds$, ${\bf x}'(s)\equiv d{\bf x}(s)/ds$,
and
${\cal H}({\bf p},p_4)=p_4^{1-\gamma}+D_\lambda(\hat{\bf p}^2)^{\lambda/2}$.
At each $s$,
the integrals over the components of
 ${\bf p}(s)$
run
from $-\infty$ to $\infty$,
whereas
those over
 $p_4(s)$
run
from $-i\infty$ to $i\infty$.
To obtain the distribution $P({\bf x},t)$, we finally form the integral
\begin{equation} \label{@PIintS}
P({\bf x},t) = \int_0 ^{\infty}ds \{ {\bf x} \, t\,s| {\bf 0}\,0 \,0\}.
\end{equation}
This is analogous to prescription (\ref{@PXT1}) which links solutions of the space- and time-fractional diffusion equations (\ref{@SF1}) and (\ref{@TF1}).

If $\gamma=0$,
the path integral over
$p_4(s) $ yields the functional
$\delta[t'(s)-1]$, which ensures that $dt$ and $ds$ increments are equal.
This brings
(\ref{@CFP})
to the canonical
path integral
\begin{eqnarray}
({\bf x}_b t_b|{\bf x}_a t_a)=\int
 {\cal D} {\bf x}
{\cal D} {\bf p}
e^ {{\cal A}'},
\label{@}\end{eqnarray}
with
\begin{eqnarray}
{\cal A}'=\int d\tau [i{\bf p} \dot {\bf x}-D_\lambda(\hat{\bf p}^2)^{\lambda/2}
].
\label{@}\end{eqnarray}
Now
$P({\bf x},t)=({\bf x} t|{\bf 0}\,0)$
satisfies
the ordinary
fractional
Fokker-Planck
equation
\begin{eqnarray}
[\hat p_4+
D_\lambda(\hat{\bf p}^2)^{\lambda/2}
]P({\bf x},t)
=\delta(t)\delta^{(D)}({\bf x})\label{eq14X},
\label{@FSE}\end{eqnarray}
which has been discussed
at length
in  recent literature \cite{LASKIN}.

\section{Langevin equations and computer simulations}
 
In the past, many nontrivial
\sch{} equations (for instance that of the $1/r$-potential)
have been solved with  path integral methods
by re-formulating them on the
pseudotime axis $s$,
that is related to the time $t$ via a {\it space-dependent
differential equation\/}
$t'(s)=f(x(t))$. This method was
invented by Duru and Kleinert
\cite{DK} to solve the path integral of the hydrogen atom, and
has recently been applied successfully
to various
Fokker-Planck equations \cite{DW,TEMP}.
The stochastic differential equation
(\ref{@SDET}), that connects pseudotime $s$ and the physical time $t$,
may be seen as a stochastic version of the Duru-Kleinert transformation
that promises to be a useful tool to study
non-Markovian systems.

\begin{figure}
\includegraphics[scale=1]{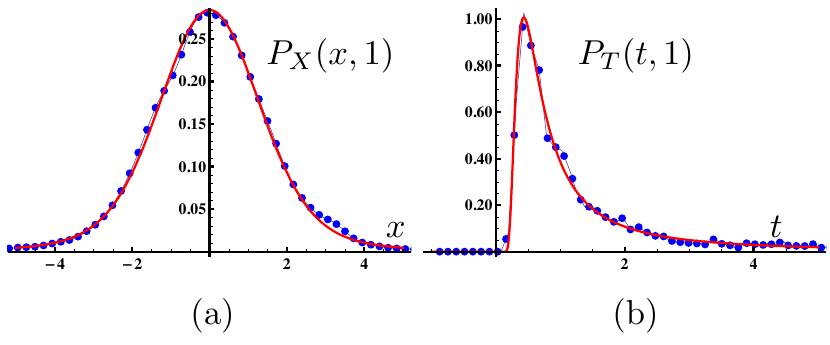}
\caption{(Color online) Comparison of analytic (solid red curve) and numerical (blue circles) results for the distribution function $P_X(x,s\!\!=\!\!1)$ in $D=1$ dimension (a), and $P_T(t,s\!\!=\!\!1)$ for $\gamma=0.3$ (b). In each case an average has been taken over $5000$ representative trajectories of stochastic differential equations (\ref{@SDEX}) and (\ref{@SDET}), with $10$ time steps $\Delta s=0.1$.}
\label{figsimPXPT}
\end{figure}

Certainly,
the solutions of Eq.~(\ref{@FSE})
can also be obtained from  a stochastic differential equation
\begin{eqnarray}
\dot{\bf x}=\etab,
\label{@SDEX}\end{eqnarray}
whose noise is distributed with a 
fractional probability
\begin{eqnarray}
P[\etab]=\int\! {\cal D}^D\!p e^{
\int\! dt(i{\bf p}\cdot 
\etabs
-D_\lambda({\bf p}^2)^{\lambda/2})
}.
\label{@43}\end{eqnarray}
Simulating this stochastic differential equation on a computer, we confirm the analytic form (\ref{@PX1}) of $P_X({\bf x},s) = P({\bf x},t)$ for $\gamma=0$. See Fig. \ref{figsimPXPT} (a).

Analogously, the solution of Eq. (\ref{@TF1}) can also be obtained from a SDE
\begin{eqnarray}
t'(s)=\eta_T(s),
\label{@SDET}\end{eqnarray}
with noise distribution
\begin{eqnarray}
P[\eta_T]=\int\! {\cal D}\!p_4 e^{
\int\! ds(p_4 \eta_T
-(p_4)^{1-\gamma})
},
\label{@}\end{eqnarray}
and compared with the result (\ref{@PT}) for $P_T(t,s)$. See Fig. \ref{figsimPXPT} (b). 

\begin{figure}
\includegraphics[scale=1]{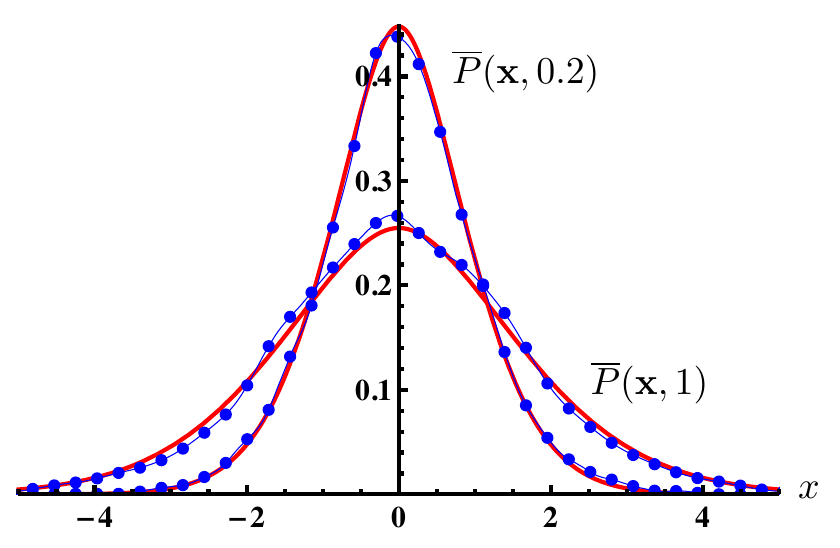}
\caption{(Color online) Comparison of computer simulation and the renormalized exact solution $\overline{P}({\bf x},t)$ for t=0.2,1.}
\label{figsimP}
\end{figure}

Solution of the double fractional Fokker-Planck equation (\ref{@10G2}) can be obtained, in view of the relation (\ref{@PIintS}) (or (\ref{@PXT1})), by simulating (\ref{@SDEX}) for $t \equiv s$ and (\ref{@SDET}), and letting the final value of the pseudotime $s$ be random. This yields a probability distribution $P({\bf x},t)$. In Fig. \ref{figsimP} we compare the results of a computer simulation with the analytic form (\ref{@11I}) by plotting $P({\bf x},t)$ as a function of ${\bf x}$ for various values of time $t$. Since the distribution $P({\bf x},t)$ itself is not normalized, but rather
\begin{equation}
\int d^D x P({\bf x},t) = \int_0^\infty \!\!\! ds P_T(t,s) = \frac{\theta(t) t^{-\gamma}}{\Gamma(1-\gamma)},
\end{equation}
we define a renormalized version $\overline{P}({\bf x},t) = P({\bf x},t)/\int d^D x P({\bf x},t)$.

\comment{
Note that there is a stochastic relation between the time $\tau$ and the
pseudotime $s$ for which the \sch{}r equation
is valid. It is found
by solving
the stochastic differential equation
\begin{eqnarray}
\tau'=\eta(s),
\label{@}\end{eqnarray}
where the noise has the distribution
\begin{eqnarray}P[\eta]
\equiv\int{\cal D} E\exp\left\{\int ds\,[
iE
\eta
-E^{1-\gamma}]
\right\}\!.
\label{@}\end{eqnarray}
On the $s$-axis, the amplitudes
are Markovian.
On the $\tau$-axis, this is no longer true.
}

\section{Summary}
Summarizing, we have seen that a many-body system with strong couplings between the constituents
satisfies
a more general form of the \sch{} equation, in which the momentum and the energy
appear with a power different from $\lambda=2$ and $\gamma=0$, respectively.
We have calculated the associated Green functions and discussed their properties and their representations.
We pointed out that these Green functions can be written as path integrals over fluctuating
 time and space orbits that are functions of some pseudotime $s$.
This is a Markovian object, but non-Markovian in the physical time
$t$. The  non-Markovian character is caused
by the fact that function $t(s)$ follows
a stochastic differential equation
of the Langevin type.

The particle distributions
can also be obtained by solving a
 Langevin type of
equation
in which the noise has correlation functions
whose probability distribution
is  specified by an equation like (\ref{@43}).

The Green functions whose theory was presented here 
will play an important role in the development 
of an interacting theory 
of fields whose 
worldlines contain non-Gaussian 
random walks
displaying extremely large deviations from their avarages.

{
{~\\
Acknowledgment:
We are grateful to P. Jizba, 
and A.~Pelster
for useful comments.
V.Z. received support from GA\v{C}R Grant No. P402/12/J077.
}}

~\\
\comment{{\bf Appendix 1}:
The lowest-order critical exponents
can be extracted directly
from the one-loop-corrected inverse Green function $G^{-1}(E,{\bf p})$
in $D=2+\epsilon$ dimensions after a
minimal subtraction of the $1/\epsilon$ -pole at \cite{ABBR}:
\begin{eqnarray}
E\!-\!{\bf p}^2\!+
a
\left(\sfrac{1}3{\bf p}^2\!-\!E\right)^{D-1}
\comment{
\left[\frac{\Gamma(1-D)}{
(\sqrt{8}\pi)^D(3!)^{D/2+1}}\!-\!\frac{1}{288 \pi^2 \epsilon}\right]
}
.
\label{@}\end{eqnarray}
For ${\bf p}=0$, this has a power
$-(-E)^{1-a\epsilon}$, so that $\gamma=a\epsilon$.
For $E=0$, on the other hand,
we obtain $(-{\bf p}^2)^{1-a\epsilon/3}$, so that
$(1-\gamma)/z-1\approx \gamma/3$.
}

{\bf Appendix 1}:
Fractional differential operators that enter the general fractional Fokker-Planck equation (\ref{@10G}) are defined through formula (\ref{@9a})
. Using $e^{-\sigma \hat{\bf p}^2} \delta({\bf x}) = (4 \pi \sigma)^{-D/2}e^{-{\bf x}^2/(4\sigma)}$, and $e^{-\sigma \hat{p}_4} \delta(t) = \delta(t-\sigma)$, we derive the following relations,
\begin{eqnarray}
|{\bf x}|^\lambda& =& \frac{\pi^{D/2} \Gamma(\frac{\lambda+D}{2})}{2^{-\lambda-D} \Gamma(-\frac{\lambda}{2})} (\hat{\bf p}^2)^{-(\lambda+D)/2} \delta^{(D)}({\bf x}) , \\
\theta(t) t^\alpha &=& \Gamma(\alpha+1) (\hat{p}_4)^{-\alpha-1} \delta(t) ,
\end{eqnarray}
which we can substitute into (\ref{@11}), (\ref{@11I}) in order to verify that these satisfy the equation (\ref{@10G2}). We first obtain
\begin{eqnarray}
P({\bf x},t) = \int_{\cal C} \frac{dz}{2\pi i} \Gamma(1+z)\Gamma(-z) D_\lambda^z (\hat{\bf p}^2)^{\lambda z/2} \nonumber\\
\times (\hat{p}_4)^{(\gamma-1)(z+1)} \delta^{(D)}({\bf x}) \delta(t) ,
\end{eqnarray}
which can be pole-expanded to yield
\begin{equation}
\sum_{n=0}^\infty (-D_\lambda)^n (\hat{\bf p}^2)^{\lambda n/2} (\hat{p}_4)^{(\gamma-1)(n+1)}
\delta^{(D)}({\bf x}) \delta(t) .
\end{equation}
Summing up this geometric series, we arrive at
\begin{eqnarray}
P({\bf x},t) = [\hat{p}_4^{1-\gamma} + D_\lambda (\hat{\bf p}^2)^{\lambda/2}]^{-1} \delta^{(D)}({\bf x}) \delta(t) .
\end{eqnarray}

{\bf Appendix 2:}
We derive several expressions for the solution $P_X({\bf x},s)$ of (\ref{@SF1}), starting from the representation (\ref{@PX1}).

On expanding the exponential, and representing the powers as $({\bf p}^2)^{\lambda n/2} = \Gamma[-\lambda n/2]^{-1} \int_{0}^\infty \frac{d\sigma}{\sigma} \sigma^{-\lambda n/2} e^{-\sigma {\bf p}^2}$, the momentum integration yields the superposition
of Gaussian expression
\begin{equation}
P_X({\bf x},s) = \int_{0}^\infty \frac{d\sigma}{\sigma} f_\lambda(\sigma) 
P_G({\bf x},D_\lambda^{2/\lambda} s^{2/\lambda} \sigma) ,\label{@52}
\end{equation}
with weight
\begin{equation}
f_\lambda(\sigma) = \sum_{n=0}^\infty \frac{(-1)^n \sigma^{-\lambda n/2}}{n! \Gamma(-\lambda n/2)}.
\end{equation}
To prove this, we
perform the $\sigma$-integration term by term, using the formula $\int_0^\infty \frac{d\sigma}{\sigma} \sigma^{-\nu} e^{-a/\sigma} = \Gamma(\nu)/a^\nu$, and
obtain the large-$|{\bf x}|$ expansion
\begin{equation} \label{@PX15} 
P_X({\bf x},s) = \frac{1}{\pi^{D/2} |{\bf x}|^D}
\sum_{n=0}^\infty \frac{(-1)^n \Gamma(\frac{\lambda n + D}{2})}{n! \Gamma(-\lambda n/2)}
\left[ \frac{\ell_s^\lambda}{|{\bf x}|^\lambda} \right]^{n} \!\!\!,
\end{equation}
where $\ell_s = 2(D_\lambda s)^{1/\lambda}$.
The series can also be viewed as a pole expansion of the contour integral, and hence
\begin{equation} \label{@PX2}
P_X({\bf x},s) = \frac{1}{\pi^{D/2} |{\bf x}|^D}
\int_{\cal C} \frac{dz}{2\pi i} \frac{\Gamma(\frac{\lambda z + D}{2}) \Gamma(-z)}{\Gamma(-\lambda z/2)}
\left[
\frac{|{\bf x}|^{\lambda}}{\ell_s^\lambda}
 \right]^{-z}\!\!\!,
\end{equation}
with the contour ${\cal C}$ running from $-i\infty$ to $+i\infty$. 
From this, 
the expansion (\ref{@PX15}) arises by enclosing the right complex half-plane and calculating the residua of the integrand, using ${\rm Res}( \Gamma(a z + b), -(n+b)/a ) = (-1)^n/(n! a)$. A small-$|{\bf x}|$ 
expansion of (\ref{@PX2}) is obtained by closing the integration contour in
the left half-plane, leading to
\begin{equation} \label{@PX3} 
P_X({\bf x},s) = 
\sum_{n=0}^\infty \frac{(-1)^n 2/\lambda}{\pi^{D/2}\ell_s^D}
 \frac{\Gamma(\frac{2n+D}{\lambda})}{n! \Gamma(\frac{D}{2}+n)}\!
\left[
\frac{|{\bf x}|^{2}}{
\ell_s^2}
 \right]^{n}.
\end{equation}
The series (\ref{@PX15}) and (\ref{@PX3}) are convergent, or asymptotic, or even trivially zero, depending on the parameter $\lambda$.


\begin{thebibliography}{9}


\bibitem{BP}
\aut{W. Feller}
{\it An Introduction to Probability Theory and Its Applications\/}
vol. 2, Wiley, New York, 1991;
\aut{J.-P. Bouchaud} and \aut{M. Potters},
 {\it Theory of Financial Risks\/},
{\it From Statistical Physics to Risk Management\/},
Cambridge U. Press,
2000. See also Ch. 20 in \cite{PI}.



\bibitem{BLS} F. Black and M. Scholes,
J. Pol. Economy {\bf 81}, 637 (1973).



\bibitem{PIR2}
The theory has been reviewed in many detailed  publications, our notation 
follows 
 the textbook \cite{PI}.
%

\bibitem{PI} H. Kleinert, {\it Path Integrals in Quantum Mechanics,
Statistics, Polymer Physics, and Financial Markets},
World Scientific, Singapore, 2006.
%

\bibitem{PREIS}
T. Preis, {\it Econophysics in a Nutshell}, Science \&  Culture 76, 333-337 (2010)

\bibitem{MANST}
R. N. Mantegna, H. E. Stanley,
{\it Introduction to Econophysics: Correlations and Complexity in Finance},
Cambridge University Press (2000)

\bibitem{PODOB}
B. Podobnik, P. Ch. Ivanov, Y. Lee, A. Chessa and H. E. Stanley, Europhys. Lett. {\bf 50} 711 (2000) 
(arXiv:cond-mat/9910433)



\bibitem{remn}
A travelling  pedestrian
salesman is a Gaussian
random walker, as a jetsetter he becomes a L\'evy random walker.


\bibitem{TAL}
See {\tt http://en.wikipedia.org/wiki/Black\_swan\_theory}


\bibitem{PIR} The concept of a Hamiltonian in the theory of statistical 
distributions was introduced in the textbook \cite{PI}. It has its root in 
the path integral formulation of quantum mechanics
and
emphasizes 
the fact that in this
formulation
particles run along fluctuating world lines in spacetime, where
they 
perform random walks 
with distributions
that may be Gaussian or nongaussian
depending on the form of the Hamiltonian 
as functions of the momentum $p$.
The distributions solve a Fokker-Planck equation
of the type
(\ref{@10G}),
 driven by differential 
operator that is obtained  by  
replacing the momentum in the Hamiltonian $H(p)$ by the differential operator $-i \partial_x$.




\bibitem{RUFFO}
C. Nardini, S. Gupta, S. Ruffo, T. Dauxois, and F. Bouchet,
	J. Stat. Mech. (2012) L01002
(http://arxiv.org/pdf/1111.6833.pdf)



\bibitem{VS} 
R. E. Angulo, V. Springel, S. D. M. White, A. Jenkins, C. M. Baugh, C. S. Frenk, 
Monthly Notices of the Royal Astronomical Society 426: 2046–2062 (2012)
(arXiv:1203.3216).



\bibitem{DU} Du Jiulin, 
 2004 Europhys. Lett. {\bf 67}, 893 (2004),
Phys. Lett. A {\bf 329}, 262(2004).


\bibitem{EI} 
J. Einasto,
(arXiv:1109.5580).


\bibitem{PEINKE}
\aut{F. Boettcher}, \aut{C. Renner}, \aut{H.P. Waldl}, \aut{J. Peinke},
Boundary-Layer Meteorology
(2003) Volume 108, Issue 1, pp 163-173
(arXiv:physics/0112063).



\comment{
\bibitem{SEISMI}
\aut{P. Bhattacharyya} and \aut{B.K. Chakrabarti} (Eds.)
{\it Modelling Critical and Catastrophic Phenomena in Geoscience: A Statistical
Physics Approach\/},
Lecture
Notes in Physics 705 (Springer, Berlin, 2006).
}
\bibitem{SEISM}
P. Bhattacharyya, A. Chatterjee, B.K. Chakrabarti,
Physica A {\bf 381}, 377 (2007)
(arXiv:physics/0510038) .


\bibitem{GMTS}
S. Umarov, C. Tsallis, M. Gell-Mann, and S. Steinberg,
J. Math. Phys. 51, 033502 (2010)
(http://arxiv.org/abs/cond-mat/0606040).



\bibitem{REM2}
Note that the function $\tilde H(\eta)$ 
is the functional Fourier transform of the Hamiltonian
(\ref{@HAM}) that drives the Fokker-Planck equation
(\ref{@10G}). 
See Eq.~(20.153) 
in Ref.~\cite{PI}.

\bibitem{ShlZas}
M.F. Shlesinger, G.M. Zaslavsky, U. Frish (Eds.), {\it L\'{e}vy Flights and Related Topics in Physics}, LNP {\bf 450} (1995).

\bibitem{PekSzn}
R. Kutner, A. Pekalski, K. Sznajd-Weron (Eds.), {\it Anomalous Diffusion From Basics to Applications}, LNP {\bf 519} (1999).

\bibitem{Sro}
T. Srokowski, Phys. Rev. E {\bf 79}, 040104(R) (2009).

\bibitem{REMMK}
Fokker-Planck equations 
in which only the time derivative
has a fractional power have been studied in 
R. Metzler and J. Klafter, Physics Reports {\bf 339} 1-77 (2000).

\bibitem{KS}
{H. Kleinert} and
V. Schulte-Frohlinde%
{V. Schulte-Frohlinde},
    {\it Critical Phenomena in $\phi ^4$-Theory\/},
    World Scientific, Singapore, 2001
({\tt http://klnrt.de/b8}).

\bibitem{HKN}
{H. Kleinert},
EPL {\bf 100}, 10001 (2012)
({\tt http://klnrt.de/399/399-TAIPEH.pdf}).
{H. Kleinert},
({\tt http://klnrt.de/403}).


\bibitem{MEER}
M. M. Meerschaert, A. Sikorskii, {\it Stochastic Models for Fractional Calculus}, Walter de Gruyter (2011).

\bibitem{ZAS}
G. M. Zaslavsky, {\it Hamiltonian Chaos and Fractional Dynamics}, OUP Oxford (2005).


\bibitem{Hil}
R. Hilfer (Ed.), {\it Applications of Fractional Calculus in Physics}, World Scientific (2000).

\bibitem{BMK}
E. Barkai, R. Metzler, and J. Klafter, Phys. Rev. E {\bf 61}, 132–138 (2000).


\bibitem{Feynman}
R.P. Feynman, Phys. Rev. {\bf 80}, 440
 (1950);
R. P. Feynman and A.R. Hibbs, \textit{Quantum Mechanics
and Path Integrals} (McGraw-Hill, New York, 1965).
%


\comment{
\bibitem{STR}
H. Kleinert, 
Phys. Rev. D {\bf 57}, 2264 (1998) (cond-mat/9801167).
%
}
%
\comment{
\bibitem{rLipa}
\comment{
{J.A. Lipa, D.R. Swanson, and J. Nissen,
T.C.P. Chui and U.E. Israelson, Phys. Rev. Lett. {\bf 76}, 944 (1996).}\\
See also: {D.R. Swanson, T.C.P. Chui, and J.A. Lipa,
Phys. Rev.~B {\bf 46}, 9043 (1992);
D. Marek, J.A. Lipa, and D. Philips, Phys. Rev.~B {\bf 38}, 4465 (1988).
Also see picture on the titlepage of the textbook \cite{KS22}.}
}
J.A. Lipa, J.A. Nissen, D. A. Stricker,
D.R. Swanson,
and
T.C.P. Chui,
Phys. Rev. B {\bf 68}, 174518 (2003);
 M. Barmatz, I. Hahn
J.A. Lipa,
R.V. Duncan,
Rev. Mod. Phys. {\bf 79}, 1 (2007).
Also see picture on the titlepage of the textbook \cite{KS22}.
\bibitem{KHE}
H. Kleinert,
Phys. Lett. A {\bf 277}, 205 (2000) (cond-mat/9906107);
Phys. Rev. D {\bf 60}, 085001 (1999) (hep-th/9812197).
%
\bibitem{MVF}
 H. Kleinert,
 {\it Multivalued Fields in Condensed Matter,
Electromagnetism, and Gravitation\/},
          World Scientific, Singapore, 2008, pp. 1--497
({\tt http://klnrt.de/b11}).
}

\comment{
\bibitem{ABAR}
V.N. Gribov and A.A. Migdal. Soviet Phys. JETP {\bf 28}, 7841  (1969);
See also
\aut{H.D. Abarbanel}, \aut{J.D. Bronzan}, \aut{R.L. Sugar}, and
\aut{A.R. White}, Phys. Reports {\bf 21}, 119 (1975)
[in particular Eq. (4.27)].
%
%
\bibitem{KS22}
\aut{H. Kleinert} and \aut{V. Schulte-Frohlinde},
{\it Critical Properties
of $\phi^4$-Theories},
World Scientific, Singapore
2001
({\tt klnrt.de/b8}).
%
%
%
\bibitem{BPZ}
In two dimensions, a sequence of
critical exponents
have been tabulated in
\aut{A.A. Belavin},
\aut{A.M. Polyakov},
 and \aut{A.B. Zamolodchikov},
Nucl. Phys. B {\bf 241}, 333 (1984).
%
\bibitem{GML}
M. Gell-Mann and F. E. Low, Phys. Rev. {\bf 95}, 1300 (1954).
}
\bibitem{METZz}

For the  so-called Riesz 
 fractional derivative
see
R. Metzler, E. Barkai, J. Klafter,
Phys. Rev. Lett. {\bf 82}, 3564 (1999);
B.J. West, P. Grigolini,
R. Metzler,
and
T.F. Nonnenmacher, Phys.Rev. E {\bf 55}, 99 (1997).
For the so-called Weyl derivative.
See
R.K. Raina and C.L. Koul, Proc. Am. Math. Soc. {\bf 73}, 188 (1979).




\bibitem{METZ}
The relevant functional matrix is
$\langle{\bf x} |
(-\nablabf^2)^{\lambda/2}|{\bf x} '\rangle=\Gamma[-\lambda/2]^{-1}
\int d\sigma\, \sigma^{-\lambda/2-1}
{(4\pi \sigma)^{-D/2}}e^{R^2
/4\sigma}
$=
$\!\!{}^D\! c_\lambda R^{-\lambda-D}$, where
${}^D\!c_\lambda=2^{\lambda}\Gamma((D+\lambda)/2)/\pi^{D/2}\Gamma(-\lambda/2),$
and $R\equiv|{\bf x}-{\bf x}'|$. If $\lambda$ is close to an
even integer, it needs a small positive shift $\lambda\rightarrow\lambda_+\equiv\lambda+\epsilon$ and we can replace $ \epsilon R^{\epsilon-1} $ by 
$\delta(R)=S_D R^{D-1}\delta^{(D)}({\bf R})$, where $S_{D}=2\pi^{D/2}/\Gamma(D/2)$.
For  $A>0$
 we have $|{\bf x}'
|^{-A}= {}^D\!c_{\lambda_A}^{-1}\langle{\bf x}' |
(-\nablabf^ 2)^{\lambda_A/2}|{\bf 0} \rangle$ with
$\lambda_A\equiv A-D$, 
so that we find
$\int d^Dx'\langle{\bf x} |(-\nablabf^ 2)^{\lambda/2}|{\bf x}' \rangle |{\bf x}'|^{-A}=
{}^D\!c_{\lambda_A}^{-1}\langle{\bf x}|(-\nablabf^ 2)^{(\lambda+A-D)/2}
|{\bf 0}\rangle =
{}^D\!c_{\lambda+A-D}{}^D\!c^{-1}_{\lambda_A}|{\bf x}
|^{-A-\lambda}$. 




\bibitem{GRAD}
I. S. Gradshteyn and I. M. Ryzhik, {\it Table of Integrals, Series, and Products, Seventh Edition}, Elsevier (2007),
Formula $3.382.7$

\bibitem{Erdelyi}
A. Erd\'{e}lyi, W. Magnus, F. Oberhettinger, and F. Tricomi, {\it Higher Transcendental Functions, Vol. 3}, New York (1981), pp. 206-212.

\bibitem{Haubold}
H. J. Haubold, A. M. Mathai, and R. K. Saxena, J. Appl. Math. 298628 (2011).

\bibitem{REMHamiltonian}
More generally, $\hat{H}$ can be a generic time-independent Hamiltonian. In particular, it may contain an additional potential term.


\bibitem{Nolan}
Nolan, J. P., {\it Multivariate stable distributions: approximation, estimation, simulation and identification}. In R. J. Adler, R. E. Feldman, and
M. S. Taqqu (Eds.), {\it  A Practical Guide to Heavy Tails}, pp. 509-526,
Birkhauser, Boston
  (1998).

\bibitem{BarSil}
E. Barkai  and R. J. Silbey, J. Phys. Chem. B, 104 (16), pp 3866–3874 (2000).

\bibitem{REMstable}
L\'{e}vy distributions are implemented in {\scshape Wolfram Mathematica 8} under the command {\it StableDistribution}.


\bibitem{DUAN}
Jun-Sheng Duan,  J. Math. Phys. {\bf 46}, 013504 (2005).



\bibitem{FH}
C. Fox,
Trans. Amer. Math. Soc. {\bf 98}, 395 (1961).


\bibitem{Mathai}
A. M. Mathai, R. K. Saxena, H. J. Haubold,
{\it The H function: Theory and Applications},
Springer, 2010.



\bibitem{Kilbas}
A. A. Kilbas and M. Saigo, 
 Journal of Applied Math. and Stoch. Anal., {\bf 12} 2  191-204 (1999).



\bibitem{REM1}
This technique is explained in Chapters~12 and 19 of Ref.~\cite{PI}.
The pseudotime $s$ resembles the so-called
 Schwinger proper time used in relativistic physics.


\bibitem{LASKIN}
\aut{N. Laskin},
(arXiv/1009.5533);
Phys.Lett. A \textbf{268}, 298 (2000);
Phys. Rev. E \textbf{62}, 3135 (2000);
(ibid.) E \textbf{66}, 056108 (2002);
Chaos \textbf{10}, 780 (2000);
Communications in Nonlinear Science and
Numerical Simulation \textbf{12}, 2 (2007).





\comment{
\bibitem{LEVY}
R Metzler, A.V. Chechkin, V.Y. Gonchar, and J. Klafter
Chaos, Solitons, and Fractals {\bf 34}, 129 (2007).
}
\comment{
\bibitem{REMAN}
There should be  no danger of confusing the fluctuating
noise variable $\eta$
in this equation  with the constant
critical exponent $\eta$ in (\ref{@PropaGAL2}).
}
\bibitem{DK}

\aut{I.H.~Duru} and
\aut{H.~Kleinert},
Phys.~Lett. B {\bf 84\/}, 30 (1979)
({\tt klnrt.de/65/65.pdf});
Fortschr.~Phys. {\bf 30\/}, 401 (1982)
({\tt klnrt.de/83/83.pdf}).
See also Chaps. 13 and 14 in \cite{PI}.

\bibitem{DW}

\aut{A. Young} and \aut{C. DeWitt-Morette}, Ann. Phys. (N.Y.) {\bf 169},
140 (1984);
\aut{H. Kleinert} and {\aut A. Pelster},
Phs. Rev. Lett. {\bf 78}, 565 (1997).

\bibitem{TEMP}

\aut{L.Z.J. Liang}, \aut{D. Lemmens},
 and \aut{J. Tempere},
Phys. Rev. E 83, 056112 (2011)
(arxiv:1101.3713).






\end{thebibliography}

\end{document}